\documentclass[fleqn,twoside]{article}
\usepackage{espcrc2}

\usepackage{graphicx}
\usepackage[figuresright]{rotating}

\newcommand{\AmS}{{\protect\the\textfont2
  A\kern-.1667em\lower.5ex\hbox{M}\kern-.125emS}}

\title{Grand Canonical Potential for a Static Quark--Anti-quark
Pair at $\mu \ne 0$}

\author{
Z. Fodor\address[BUW]{Department of Physics, University of Wuppertal,
        Gaussstrasse 20, D-42097  Wuppertal, Germany} 
\address{Institute for Theoretical Physics, E\"otv\"os University,
P\'azm\'any P. 1/A, H-1117 Budapest, Hungary},
S. D. Katz\addressmark[BUW]\thanks{On leave from E\"otv\"os University.}, 
K. K. Szabo\addressmark[BUW] and 
A. I. T\'oth\addressmark[BUW]
}
       
\begin{document}

\vspace{-0.3cm}
\begin{abstract}
We present numerical results on the static quark--anti-quark grand 
canonical potential in full QCD at non-vanishing temperature ($T$) 
and quark chemical potential ($\mu$). Non-zero $\mu$-s are reached by means 
of multi-parameter reweighting. The dynamical staggered simulations 
were carried out for $n_f=2+1$ flavors with physical quark masses
on $4\times 12^3$ lattices.
\vspace{-0.3cm}
\end{abstract}

\maketitle

\vspace{-0.3cm}
\section{INTRODUCTION}
Forces among infinitely heavy quarks separated by distance $r$ are 
of great significance both at $T=0$ and when they are surrounded by 
an interacting medium with temperature $T$. The lattice formulation 
of their free energy was successfully carried out already 20 
years ago \cite{mcl81}. Numerous lattice results are available in quenched 
\cite{kac99} and also in full QCD \cite{kac04,petr04} at $T>0$. 

There are recent calculations to determine the free energy in channels 
with definite color transformation properties. The gauge invariance 
and the interpretation of these singlet/octet channels in view of 
\cite{bel04,phi04} raised some new, interesting questions.

All the previous results were obtained at vanishing baryonic chemical 
potential. Lattice simulations at $\mu \ne 0$ are difficult, because 
they are hindered by the sign problem of the fermion determinant. Recently, 
new techniques have been proposed to study lattice QCD at non-vanishing $\mu$
\cite{fod01,all02,for02,del03}. In this paper we use the multi-parameter 
reweighting to reach non-zero $\mu$ values. We determine the grand 
canonical potential of the heavy quark--anti-quark system up to $\mu=140$ 
MeV in quark chemical potential on $N_t=4$ lattices for physical quark masses. 
We use the same configurations as have been used in \cite{fod04}.
\vspace{-0.3cm}
\section{FORMALISM}
The free energy of a QCD system at $\mu=0$ with a static quark and 
anti-quark separated by distance $r$ can be expressed as follows
\cite{mcl81}:
\begin{eqnarray}\label{eq:free_E}
\exp(-F_{q{\bar q}}(T,r)/T+C)=\langle{L(r)L^{\dagger}(0)\rangle_T}
\end{eqnarray}
where $L(r)=\frac{1}{3}{\rm tr}\prod_{\tau=0}^{N_{\tau}-1}U_4(\tau,r)$ is the 
trace of the temporal Wilson line, i.e. the Polyakov loop and the $U_4(\tau,r)$-s 
are the link matrices in the time direction. $C$ is needed for 
renormalization, and depends only on the lattice spacing (see below). 
$F_{q\bar{q}}$ is the additional free energy due to the presence of the 
heavy quark--anti-quark pair. 

We can ask for the grand canonical potential of the heavy quark system, if we
place it into a medium with $T$ temperature and $\mu$
chemical potential. One can modify the above formula in an obvious way 
\begin{eqnarray}\label{eq:free_G}
\exp(-\Phi_{q{\bar q}}(T,\mu,r)/T+C)=\langle{L(r)L^{\dagger}(0)\rangle_{T,\mu}},
\end{eqnarray}
where $\Phi_{q{\bar q}}$ is the additional grand canonical potential due to
the presence of the heavy quark--anti-quark.

Both $F_{q\bar{q}}$ and $\Phi_{q{\bar q}}$ contain the self-energy of the 
static quarks. At $T=0$, $\mu=0$ one 
usually normalizes the potential at a given
distance: e.g. by demanding that $V_{q\bar{q}}(r_0)=0$. Here we used the
Sommer scale ($r_0=0.49$ fm). The $r$ independent shift, which is applied 
to the potential to satisfy the renormalization condition, is just the
self-energy at a given lattice spacing. This $T=0$ shift has to be removed
from the finite $T,\mu$ potentials as well. Using these renormalized
grand canonical potentials the renormalization of the Polyakov-loop can also 
be done (e.g. \cite{Kaczmarek:2002mc}).

In eq. (\ref{eq:free_G}) we need the expectation value of the Polyakov-loop 
correlator for non-vanishing $\mu$ values. For moderate chemical potentials 
one can use the multi-parameter reweighting: one generates configurations at 
$\mu=0$ and reweights the generated ensemble in the $\beta$ and $\mu$
parameters, simultaneously. The new parameters are usually chosen to improve
the overlap between the simulated and the target configurations, or saying it in
another way to reduce the systematic errors of reweighting to as small as possible. 
Starting from the transition point at $\mu=0$ the overlap is maximized along the 
transition line. However one is not forced to follow the transition line
during the reweighting. The transition ensemble contains enough information
about both the confined and the deconfined phases so a constant $T$ reweighting 
is also possible. (Though this reweighting cannot reach as far in $\mu$ as
the one along the transition line.)
The reweighting formula for the Polyakov-loop 
correlator is:
\begin{eqnarray}
\langle L(r)L^{\dagger}(0) \rangle_{T,\mu} = \frac{\langle L(r)L^{\dagger}(0) w (T,\mu)\rangle_{T_c,0}}{\langle w(T,\mu) \rangle_{T_c,0}}
\end{eqnarray}
After diagonalizing the transformed fermion matrix the $w$ weights can be 
calculated for arbitrary $\mu$ values using the explicit formula of \cite{fod01b}.

\section{RESULTS AT FINITE $\mu$}
We studied QCD with $2+1$ flavors of dynamical staggered quarks at
physical quark masses, that is at $m_{u,d}=0.0096$ and $m_s=0.25$ on $4\times
12^3$ lattices. 150 000 configurations were simulated at the critical 
gauge coupling: $\beta_{c}=5.1893$. The scale was set using the Sommer prescription.
The Wilson-loops were measured on $24\times 12^3$ lattices for relatively 
high quark masses. The scale for the physical quark mass is obtained from a
chiral extrapolation: $r_0/a=1.77(2)$. Note that the Sommer-scale  
depends weakly on the quark mass, therefore the extrapolation is quite safe. 
 
\begin{figure}[h!]
\includegraphics[width=80mm]{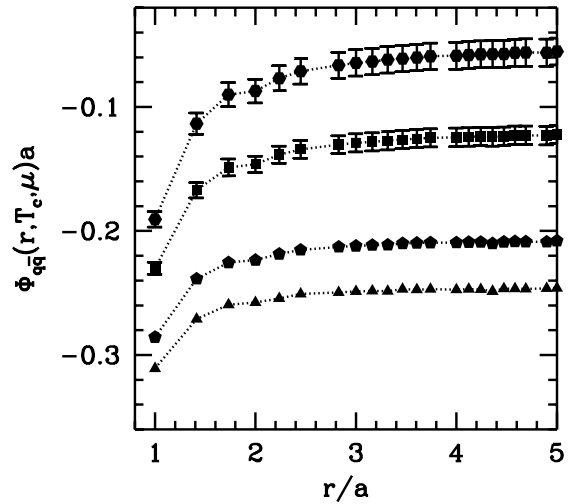}
\vspace{-1.5cm}
\caption{ 
$\Phi_{q{\bar q}}$ along the $T=T_c(\mu=0)$
line for $\mu a=0,0.05,0.10,0.15$. Higher $\mu$ corresponds
to lower curve.}
\label{fig:F_crit}
\vspace{-0.5cm}
\end{figure}

\begin{figure}[h!]
\includegraphics[width=80mm]{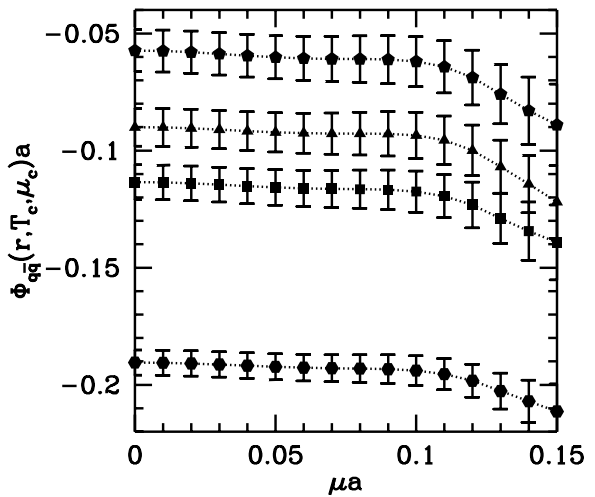}
\vspace{-1.5cm}
\caption{ 
The $\Phi_{q{\bar q}}$ as the function of $\mu$ for 
distances $(r/a)^2=1,2,3,20$ along the transition line. 
Larger $r$ corresponds to higher curve.} 
\label{fig:G_crit}
\vspace{-0.5cm}
\end{figure}

Since $\beta$ changes along the critical line, it is necessary to deal with
the problem of renormalization. In order to follow our renormalization
prescription described in sec. 2 we calculated the $T=0$ potentials not only 
at $\beta_c$ but at further two $\beta$ values. From the values of the 
potentials at a reference distance (in our case $r_0$) one
gets a $\beta$ dependent shift, which is used 
to renormalize the potentials along the critical line.

On Fig. \ref{fig:F_crit} the potential is plotted as the function of the distance 
for various chemical potentials. The reweighting was done along a constant
temperature line. For higher $\mu$ values the potential flattens out at short 
distances, whereas for smaller $\mu$ values it reaches its asymptotic value
only at larger distances.

Fig. \ref{fig:G_crit} shows the potentials as the function of $\mu$. The
reweighting is done along the transition line. Until the critical point 
(which is around $\mu a\approx 0.18$) only slight changes can be observed.  
After this point the errors get enlarged. Since there is only a small change 
in $\beta$, the renormalization has only a small effect on the result.

\begin{figure}[h!]
\includegraphics[width=80mm]{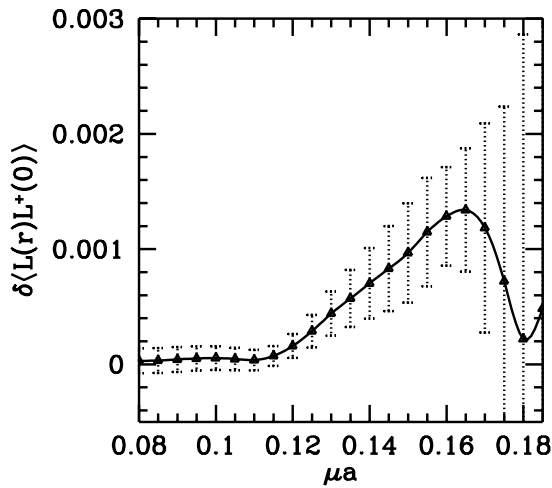}
\vspace{-1.5cm}
\caption{ 
$\delta \langle L(r) L^{\dagger}(0) \rangle =
\langle L(r) L^{\dagger}(0) \rangle _{\mu} -
\langle L(r) L^{\dagger}(0) \rangle _{\mu=0}$, i.e.
the difference in the correlator to the $\mu=0$ case as the
function of $\mu$. The distance is $(r/a)^2=13$.
} 
\label{fig:corr}
\vspace{-0.5cm}
\end{figure}
It can be instructive to examine the change in the Polyakov-loop correlator 
along the transition line. Fig. \ref{fig:corr} shows the difference in the
correlator to the $\mu=0$ case for a given distance. The correlator starts to
grow intensively before the second order endpoint. Unfortunately, at the same 
time the statistical errors get enlarged.

\section{SUMMARY}
We have determined the heavy quark--anti-quark grand canonical potentials
for non-vanishing $\mu$ values on $N_t=4$ lattices. We have presented results at fixed $T$
and various $\mu$ values as a function of $r$. We have found no significant 
change in the potential along the transition line. Around the critical endpoint we
have observed a rise in the correlations. The renormalization at
$T$=0/$T\neq$0 and $\mu$=0/$\mu\neq$0 was also discussed.\\
{\bf Acknowledgments:}
This work was partially supported by Hungarian Scientific
grants OTKA-T37615/\-T34980/\-T29803/\-TS44839/\-T46925. 
The simulations were carried out on the  
E\"otv\"os Univ., Inst. Theor. Phys. 330 P4 node parallel PC cluster.

\end{document}